\documentclass{elsart}
\usepackage{natbib}
\usepackage[dvips]{graphicx,color}

\newcommand {\hess} {\small H$\cdot$E$\cdot$S$\cdot$S$\cdot$} 

\begin{document}
\runauthor{Jim}
\begin{frontmatter}
\title{The Status of the H$\cdot$E$\cdot$S$\cdot$S$\cdot$ Project}

\author[MPI]{J. A. Hinton}
\author[All]{for the H$\cdot$E$\cdot$S$\cdot$S$\cdot$ collaboration}

\address[MPI]{Max-Planck-Institut f\"ur Kernphysik, PO Box 103980, 
D-69029 Heidelberg, Germany}
\address[All]{\tt http://www.mpi-hd.mpg.de/HESS/collaboration}

\begin{abstract}

The High Energy Stereoscopic System (\hess) - is a system of four, 
107~m$^{2}$ mirror area, imaging Cherenkov telescopes under construction 
in the Khomas Highland of Namibia (1800~m asl). The \hess~ system is 
characterised by a low threshold ($\sim$ 100~GeV) and a $\sim$1\% Crab 
flux sensitivity resulting
from the good angular resolution and background rejection provided by the 
stereoscopic technique.

The first two telescopes are operational and first results are reported here.
The remaining two telescopes (of \hess~Phase-I) will be commissioned early
in 2004. 

\end{abstract}
\end{frontmatter}

\section{System Design}

\begin{figure}
\begin{center} 
\includegraphics[height=19pc]{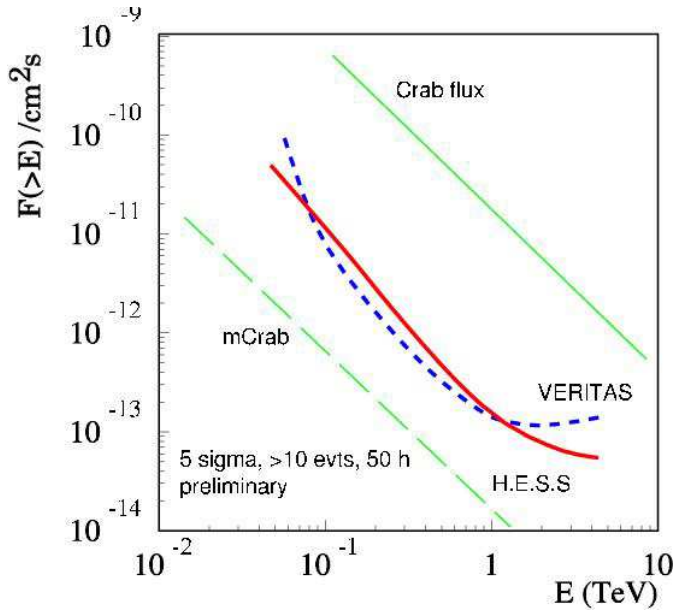}
\end{center}
\caption{Expected point-source sensitivity of \hess~Phase I compared to that of the 
VERITAS array\cite{VERITAS} (reproduced from \cite{HESS_ICRC01}).}
\label{fig_mc}
\end{figure}

\hess~is an array of 4 identical telescopes arranged in a square with 120~m sides.
Each telescope has a focal length of 15 m and a 13~m diameter. The reflectors
comprise 380 quartz-coated round facets (60~cm diameter), arranged with 
Davies-Cotton optics.

The \hess~cameras consist of 960 pixels of $0.16^\circ$ angular size 
providing a total field of view of $5^\circ$. A pixel consists of a photomultiplier
tube (PMT) with a Winston cone light collector.
The PMTs (Photonis XP2960) are organised into {\em drawers} of 16 PMTs with associated 
read-out electronics. All triggering and read-out electronics are contained 
inside the camera body.

Monte-Carlo simulations predict a sensitivity
for the 4 telescope system of around 1\% of the Crab flux 
(5$\sigma$, 50 hours) and an energy threshold of $\approx$100 GeV
(see Fig.~\ref{fig_mc}).

\section{Construction Status}

\begin{figure}[t]
\begin{center} 
\includegraphics[height=16pc]{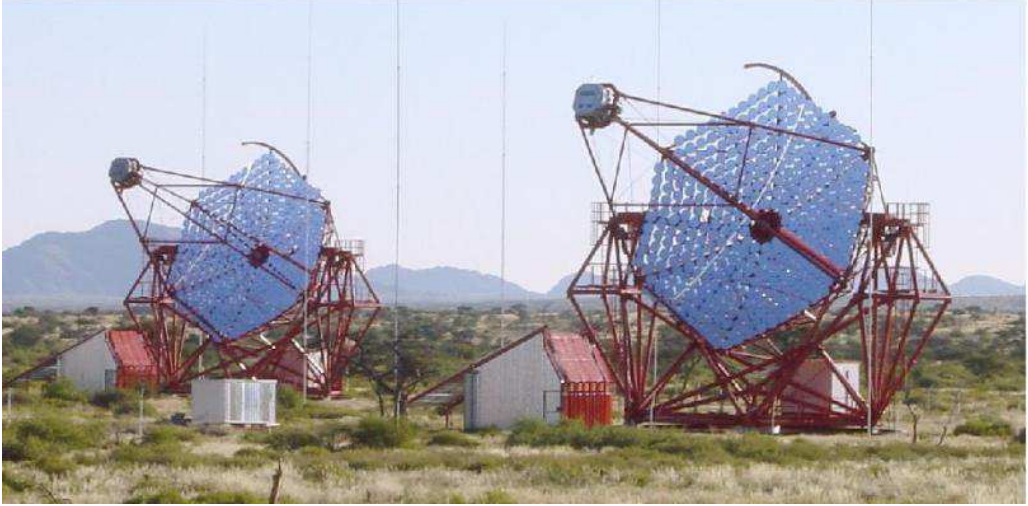}
\end{center}
\label{fig_tels}
\caption{The first two \hess~telescopes (operational March 2003).} 
\end{figure}

The first telescope of \hess~(referred to as CT3 for historical
reasons) was completed in June 2002 and immediately began Cherenkov
observations. The second telescope completed (CT2) began operations in
March 2003. Some improvements were made to the second camera (and 
upgrades to the first), 
primarily a reduction in digitisation and readout
time (from 1.6~ms to 0.6~ms) (Fig.~\ref{fig_tels}).

As of April 2003, the steel work and drive systems are complete on the
two remaining telescopes. The camera shelters, lightening masts,
residence and control buildings are all complete.  The remaining work
is the installation of mirrors on CT1 and CT4 and the installation and
commissioning of the cameras on these telescopes. The central trigger
system for the array will be installed in June 2003. All mirror facets
and PMTs for the 4 telescope system are in hand and the two remaining
cameras are mechanically complete and undergoing electronics
installation and testing in Paris.  Completion of the construction of
Phase-I is expected in early 2004.

\begin{figure}
  \begin{center}
    \includegraphics[height=14.pc]{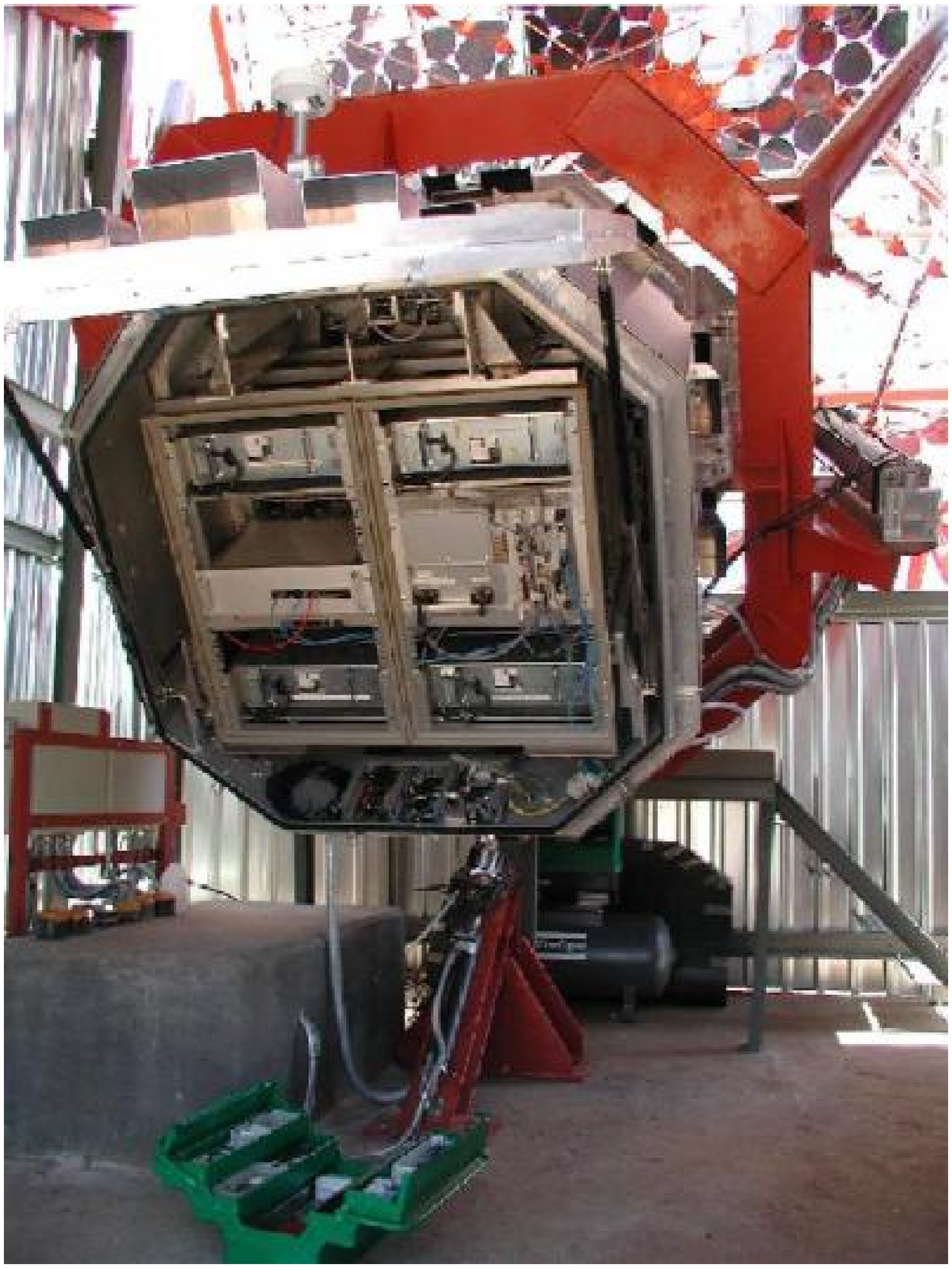}
    \includegraphics[height=14.pc]{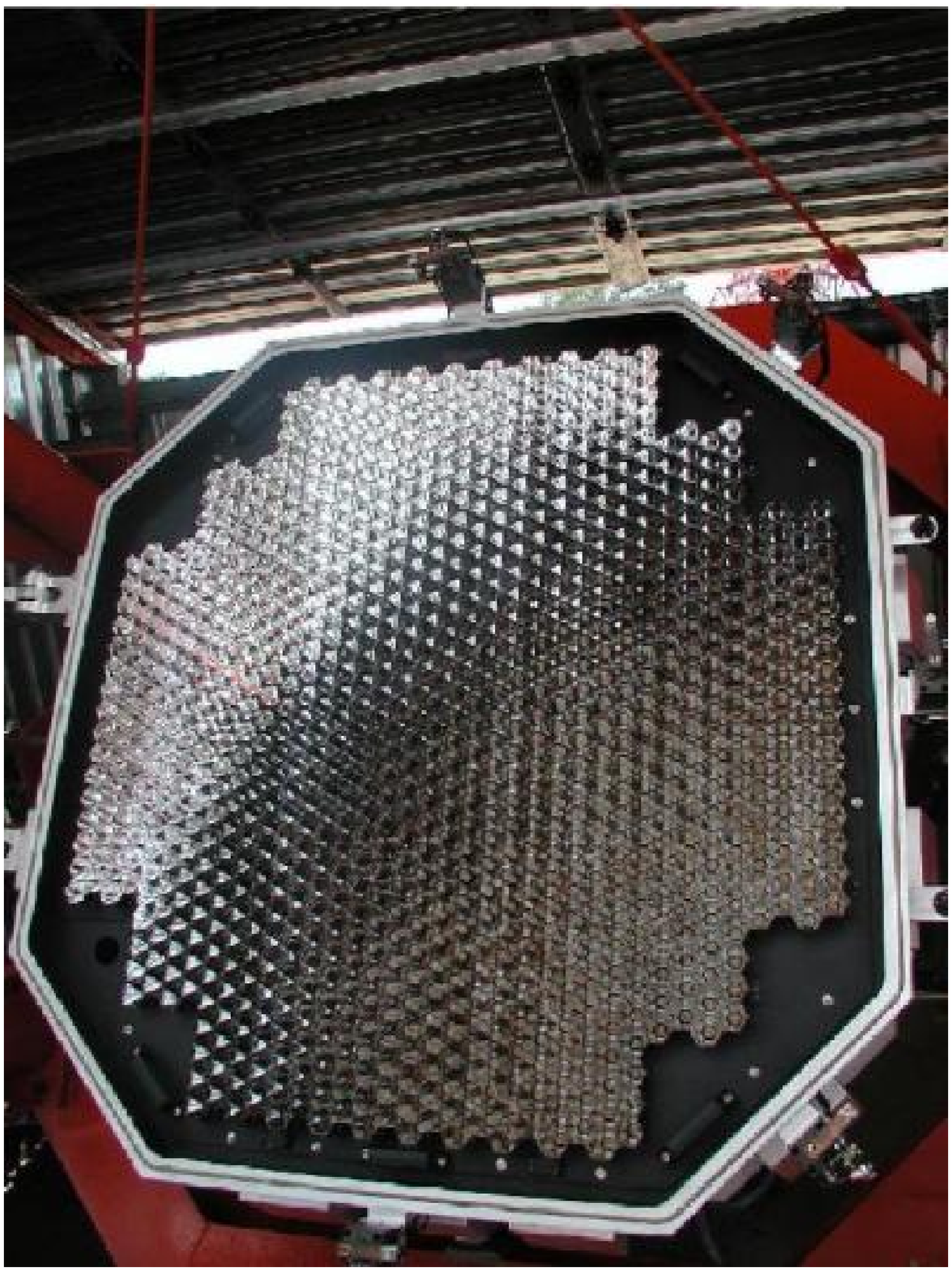}
    \includegraphics[height=14.pc]{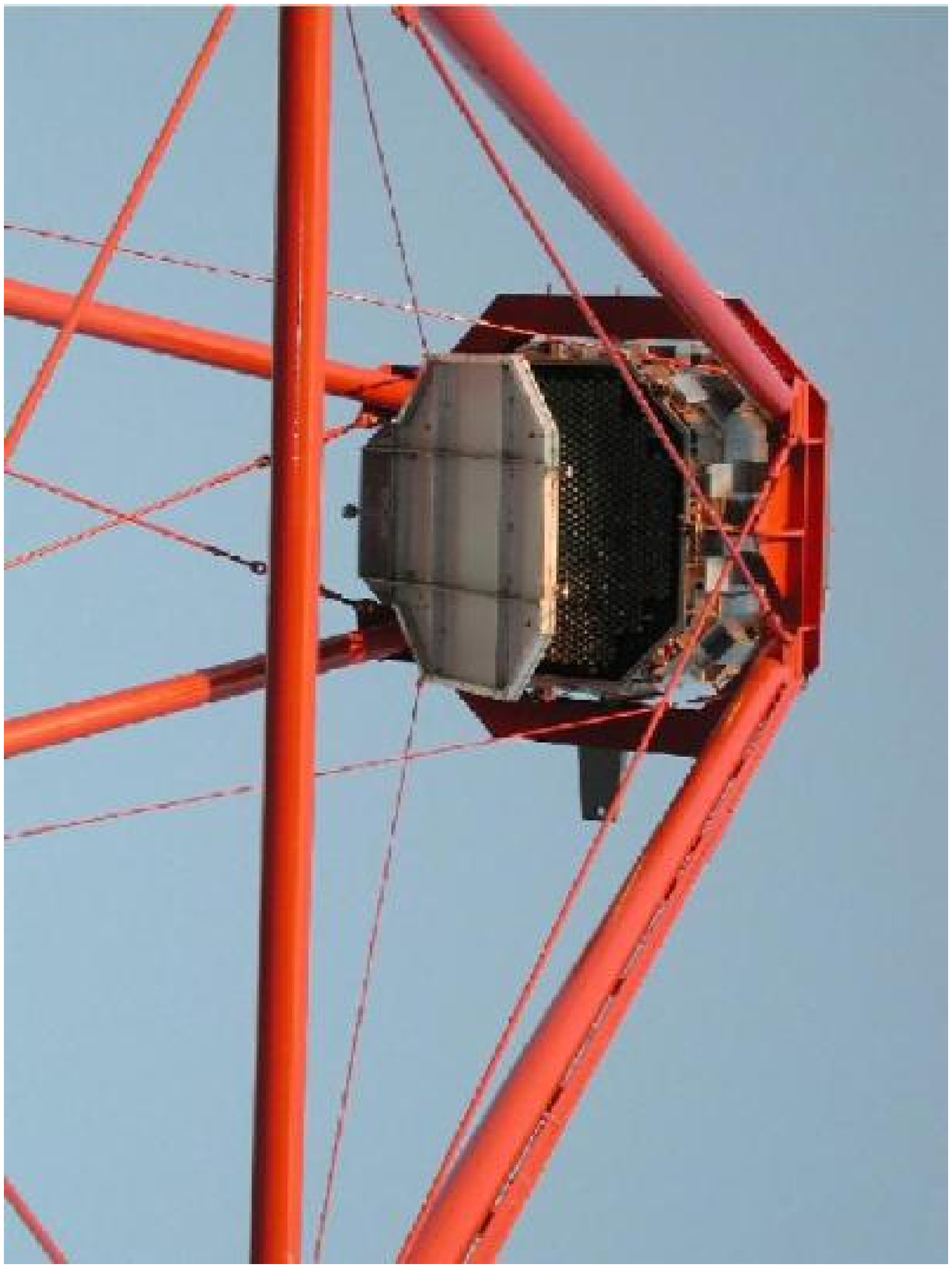}
  \end{center}
  \caption{View of the second \hess~camera. 
Left Figure: the camera readout electronics seen from the rear. 
Central Figure: the front of the camera with lid open showing the 
PMTs and Winston cones. Right Figure:  with the telescope tracking.}
\end{figure}

\section{Telescope Performance}

The optical support structure and positioning system of the \hess~telescopes
have been designed for a high level of mechanical rigidity -- we therefore expect 
to achieve good telescope tracking precision
which can be followed by small off-line corrections based, for example, on CCD
measurements. The telescope drive systems are 
equipped with encoders for control and monitoring. The {\small RMS} deviation from nominal
encoder values (over a period of 9 months) was of the order 1~arcsecond. The
slewing/positioning speed of the telescopes is $100^\circ$/minute. 

To allow remote alignment, each mirror is equipped with two
alignment motors. The alignment procedure uses the image
of a star on the closed lid of the PMT camera, viewed by a CCD camera at the 
centre of the dish. The optical point spread function of the telescope
is measured in the same way. These procedures are described in detail in~\cite{HESS_O2}.

The absolute telescope pointing accuracy has been measured using images of stars on the
camera lid. Uncorrected star images deviated from nominal position with an {\small RMS} 
error of 28''. Using a 12-parameter model to correct for misalignments of the telescopes 
axes etc., a pointing precision of 8'' is reached. Finally, 
using a guide telescope
attached to the dish for online corrections, the pointing error can be 
reduced to to 2.5'' {\small RMS}. 
\hess should therefore be able to locate bright $\gamma$-ray sources to a few arc-seconds.

\begin{figure}
\parbox{0.55\linewidth} {\centering\includegraphics[height=14pc]{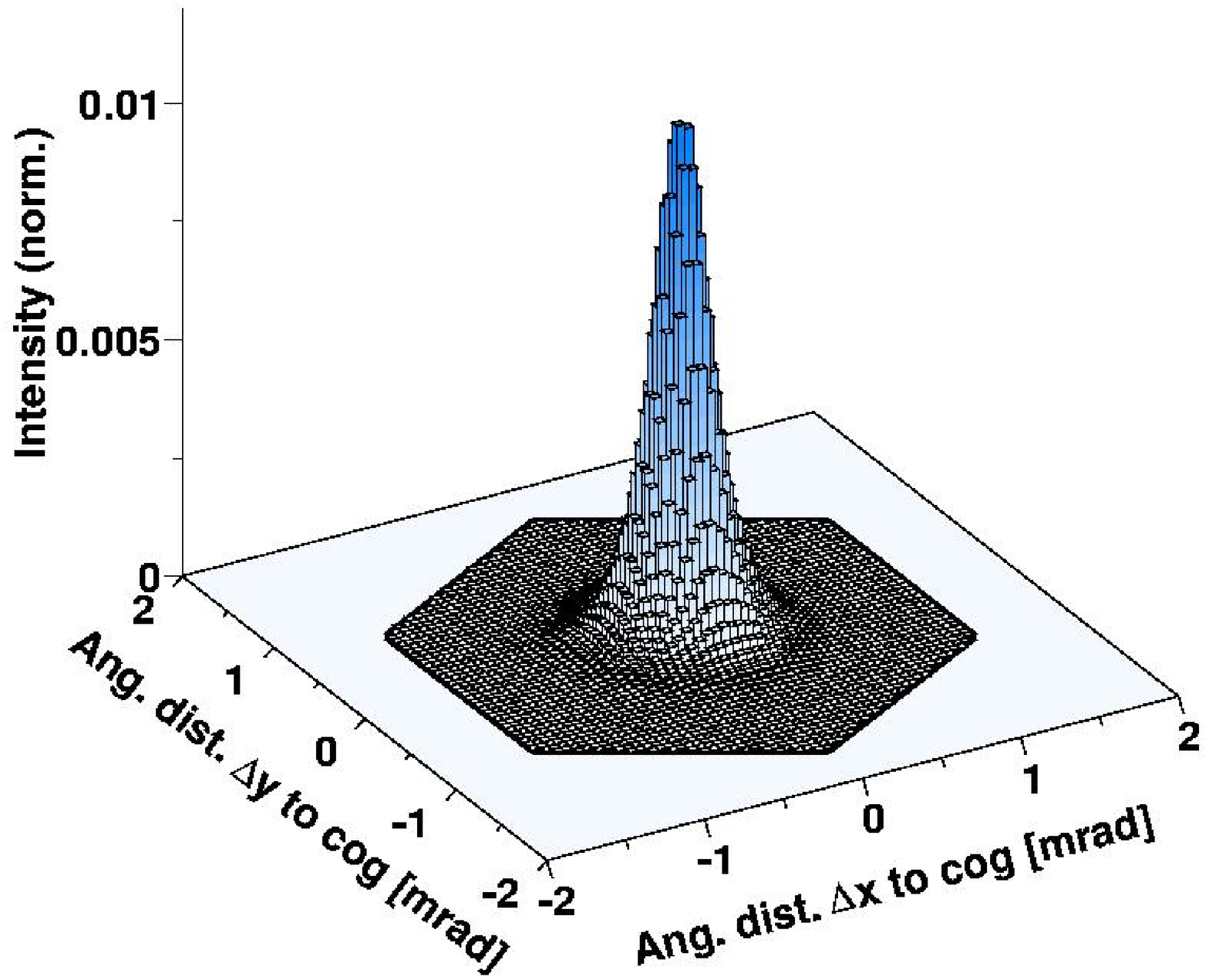}}
\parbox{0.45\linewidth} {\centering\includegraphics[height=13pc]{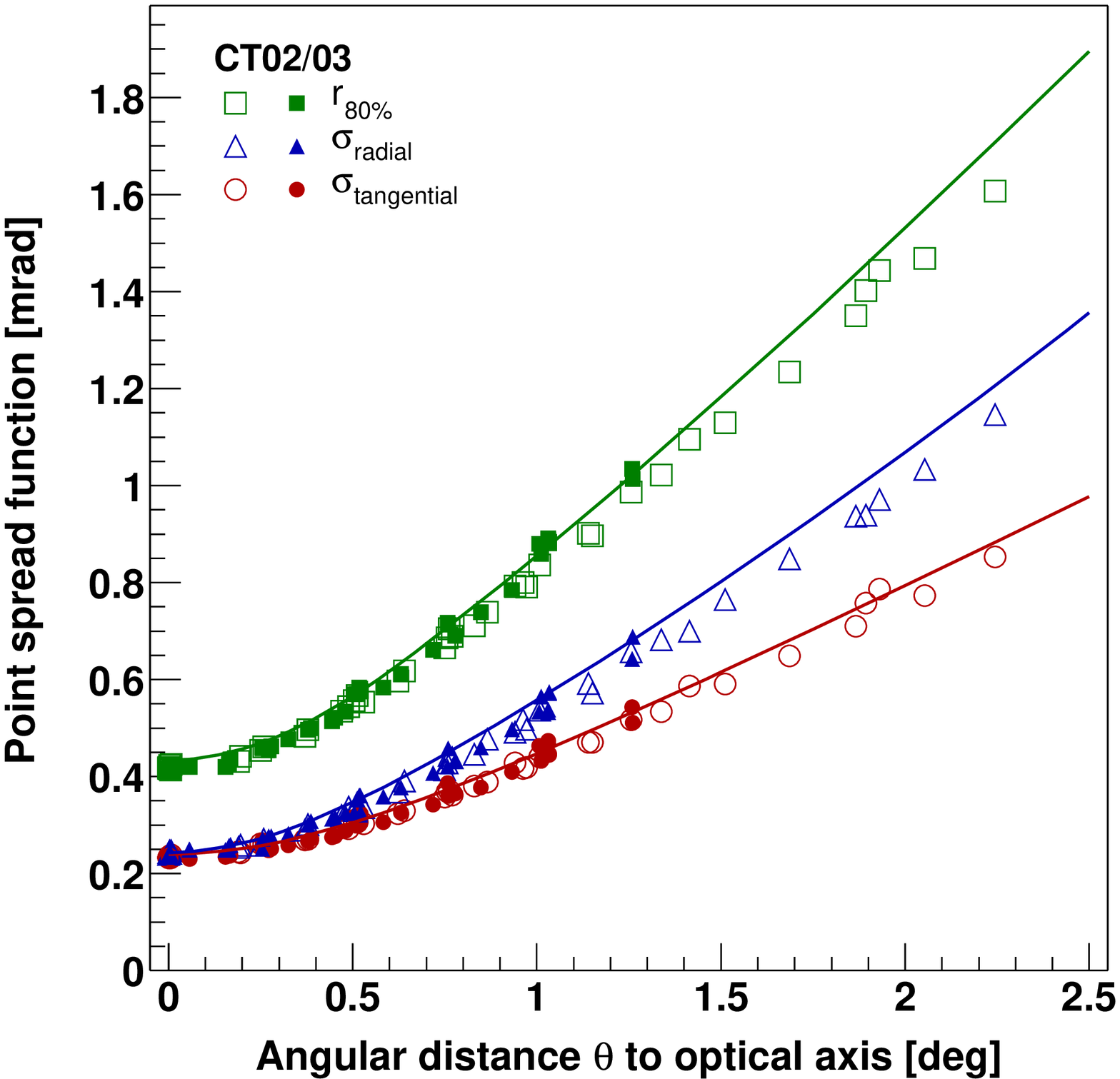}}  
\caption{Left: The on-axis point-spread function of CT3 compared 
with the pixel size (hexagon). 
Right: the behaviour of the PSF off-axis for telescopes CT2 \& CT3 
(solid lines show Monte-Carlo predictions).} 
\label{fig_psf}
\end{figure}

Figure~\ref{fig_psf} shows the optical point spread function (PSF) for the 
\hess~telescopes. The 80\% light 
containment radius is smaller than half the pixel diameter up to $2^{\circ}$ off-axis.
Measurements of the PSF over several months show no evidence for degradation since the 
installation of the camera on CT3, without need for mirror alignment.
The design and performance of the optical system of \hess~is described in detail in \cite{HESS_O1} 
and \cite{HESS_O2}.

\section{Optical Crab Pulsar}

A prerequisite of deep $\gamma$-ray observations of pulsed $\gamma$-ray 
emission from pulsars is to verify the
absolute timing system and phase reconstruction algorithm used.
The observation of optical pulsations from the Crab has been used by several
groups to provide this verification (see for example~\cite{WHIPPLE,CELESTE}). 

Optical pulsar observations have been made using a custom-built detector 
installed on the closed lid of the first Cherenkov camera (CT3). This 
instrument comprises a single PMT with current sampling every 50~$\mu$s and 
analog response time of $\approx$150~$\mu$s. Each sample is accompanied by a 
timestamp derived from the the GPS clock of the \hess~central trigger system.

\begin{figure}[t]
\parbox{0.57\linewidth} {\centering     \includegraphics[height=12pc]{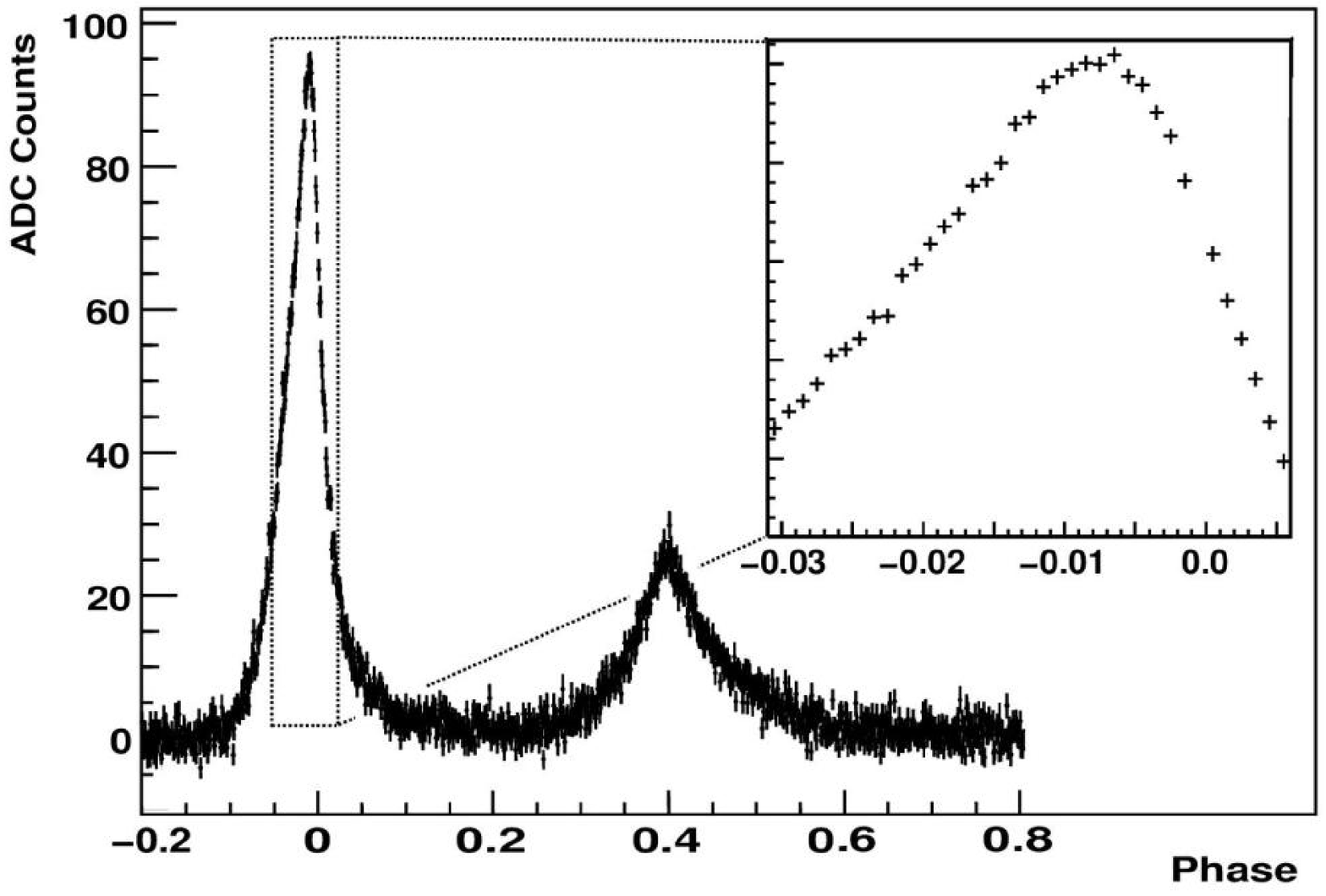}}
\parbox{0.43\linewidth} {\centering     \includegraphics[height=11pc]{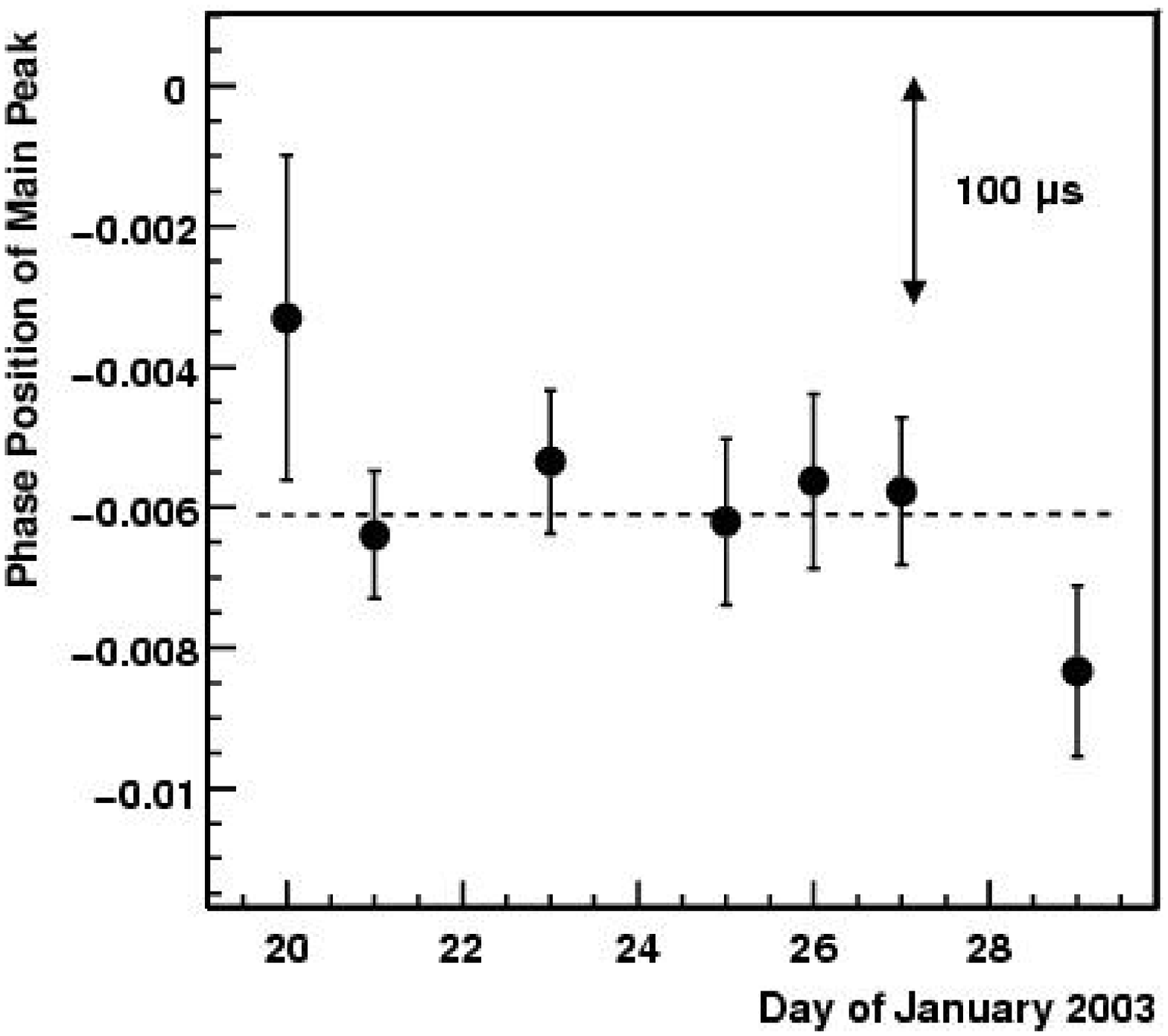}}  
\caption{Left: the Crab pulsar phasogram extracted from the full data-set of 3~hours
of best quality observations.
Right: The position of the main peak as a function of time during the measurement. 
The mean position is consistent with previous measurements
\cite{OPTIMA}.} 
\label{fig_phasogram}
\end{figure}

The large mirror area and narrow PSF of the
\hess~telescopes allows a very high signal/noise measurement of the 
pulsar light curve to be made.  Fig.~\ref{fig_phasogram} shows the light curve derived
from 3~hours of observations spread over one week in January 2003. Fig.~\ref{fig_phasogram}b
shows the mean optical flux versus phase (relative to the peak of the primary radio pulse). 

\section{Camera Performance}

Since June 2002 the first camera of the \hess~system has been taking Cherenkov
data -- operating at typical trigger rates of $\sim150$~Hz (at 30$^{\circ}$ zenith,
4 pixels $>$5~photoelectrons). The trigger rate of the telescope 
is seen to vary smoothly with threshold. At thresholds as low as 4 pixels 
$>$4~photoelectrons
there is no significant contribution of night sky background triggers to the 
rate. The observed night sky background level (from PMT currents and pedestal widths) 
is close to 100 MHz in dark regions, as expected.

Measurements of the PMT gain and flat-field coefficients (derived from laser and LED
calibration systems) show variations on the 10\% level over the nine months of operation.
After flat-fielding the performance of the camera is measured using muon-ring
images which provide the best measurement of the detector response to Cherenkov light.

\begin{figure}[b]
\parbox{0.5\linewidth} {\centering \includegraphics[height=14pc]{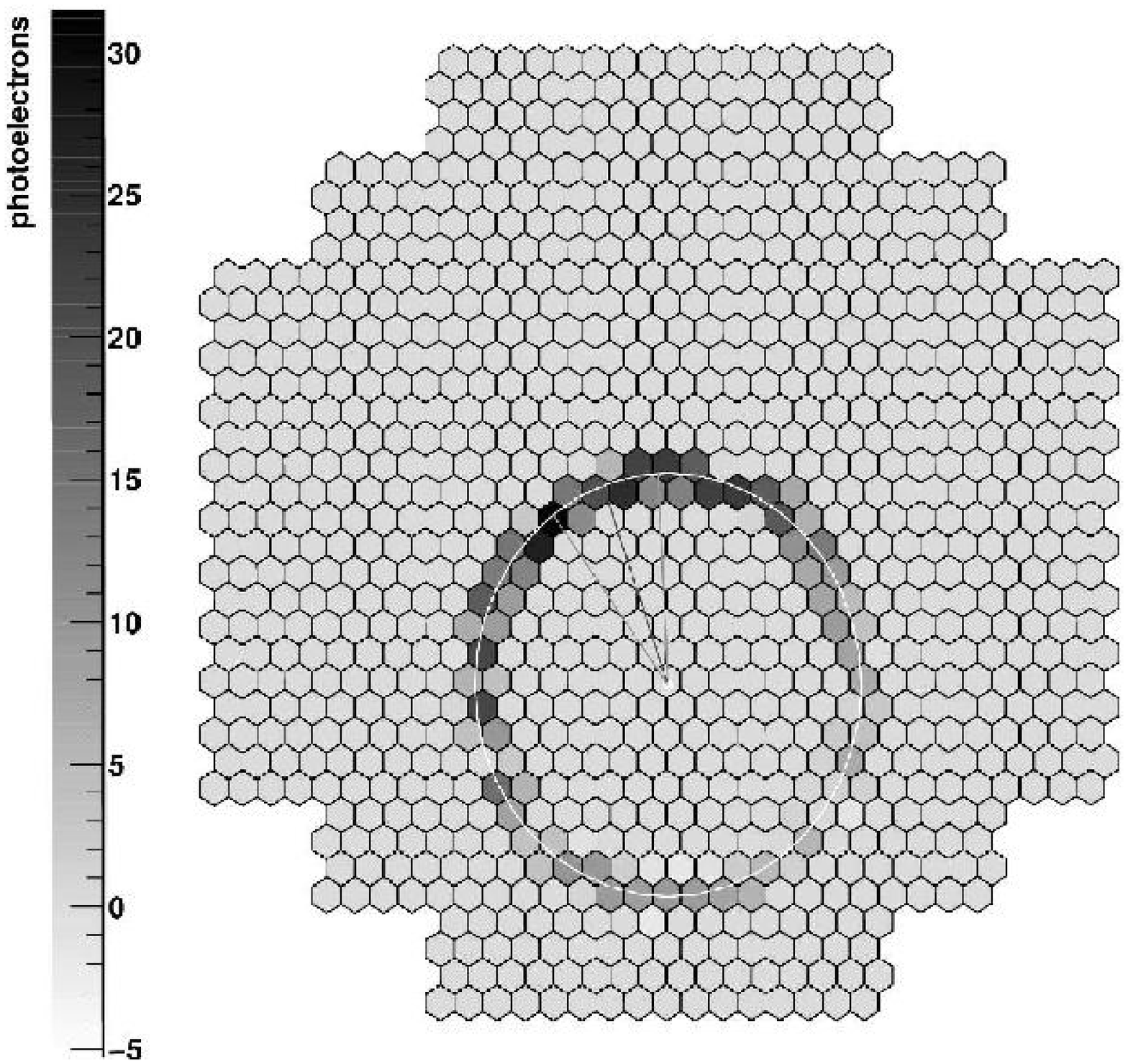}}
\parbox{0.5\linewidth} {\centering \includegraphics[height=14pc]{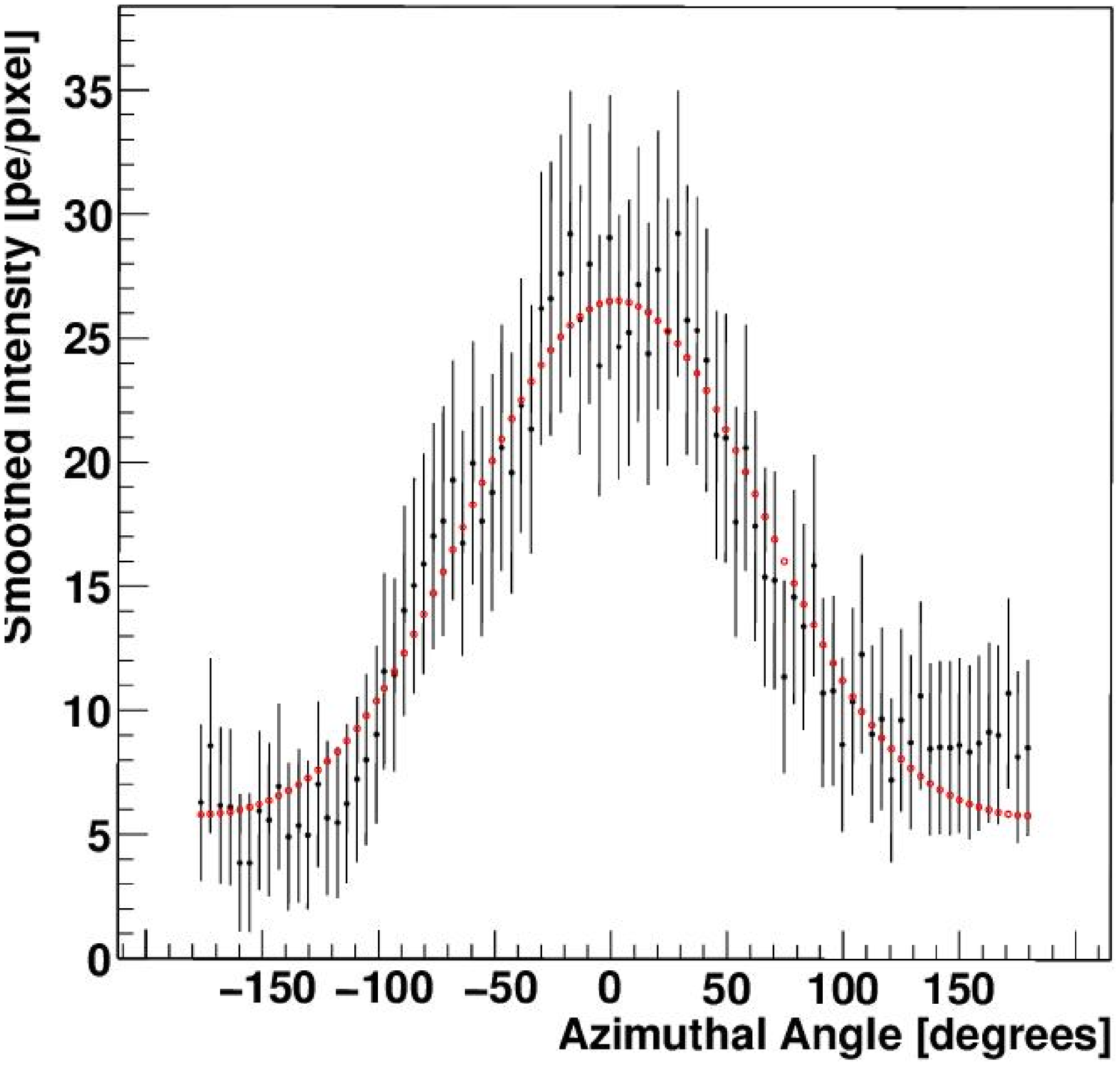}} 
\caption{Left: a cleaned Muon-ring image with the circular fit shown. 
Right: the corresponding azimuthal distribution of intensity around the ring.
The peak intensity is proportional to the detector efficiency.} 
\label{fig_muon}
\end{figure}

\vspace{1cm} Single muon images provide a powerful tool for absolute calibration
of Cherenkov Telescopes. The expected images are readily calculable 
and easily recognisable in the data (as ring-like images). 
The overall brightness of the rings
(or the normalisation of the intensity profile azimuthally around the ring) 
provides a measurement of the through-put of the system, including 
attenuation in the lower $\sim$500~m of the atmosphere, mirror reflectivity,
Winston cone collection efficiency, and PMT quantum efficiency.

Figure~\ref{fig_muon} shows a sample muon-ring image and the azimuth distribution
of intensity around the ring. The overall system efficiency can be deduced from 
a single ring with $\approx$10\% precision; such rings occur in our data at
$\sim 1$ Hz. 
Averaging over muon rings in many runs the through-put of the first telescope
seems to be $\approx$~10\% less than predicted by the simulations - within 
the systematic errors in the calculation.

Run-to-run fluctuations in the through-put deduced from muon images are at the
few percent level.

\section{Gamma-Ray Observations}

Between July 2002 and March 2003 many target objects have been observed using 
the first \hess~telescope. The list of targets is primarily composed of
objects claimed as sources by the Durham~\cite{DURHAM_INST} and 
CANGAROO~\cite{CANGAROO} groups.

\begin{figure}[h]
\parbox{0.58\linewidth} {\centering \includegraphics[width=18pc]{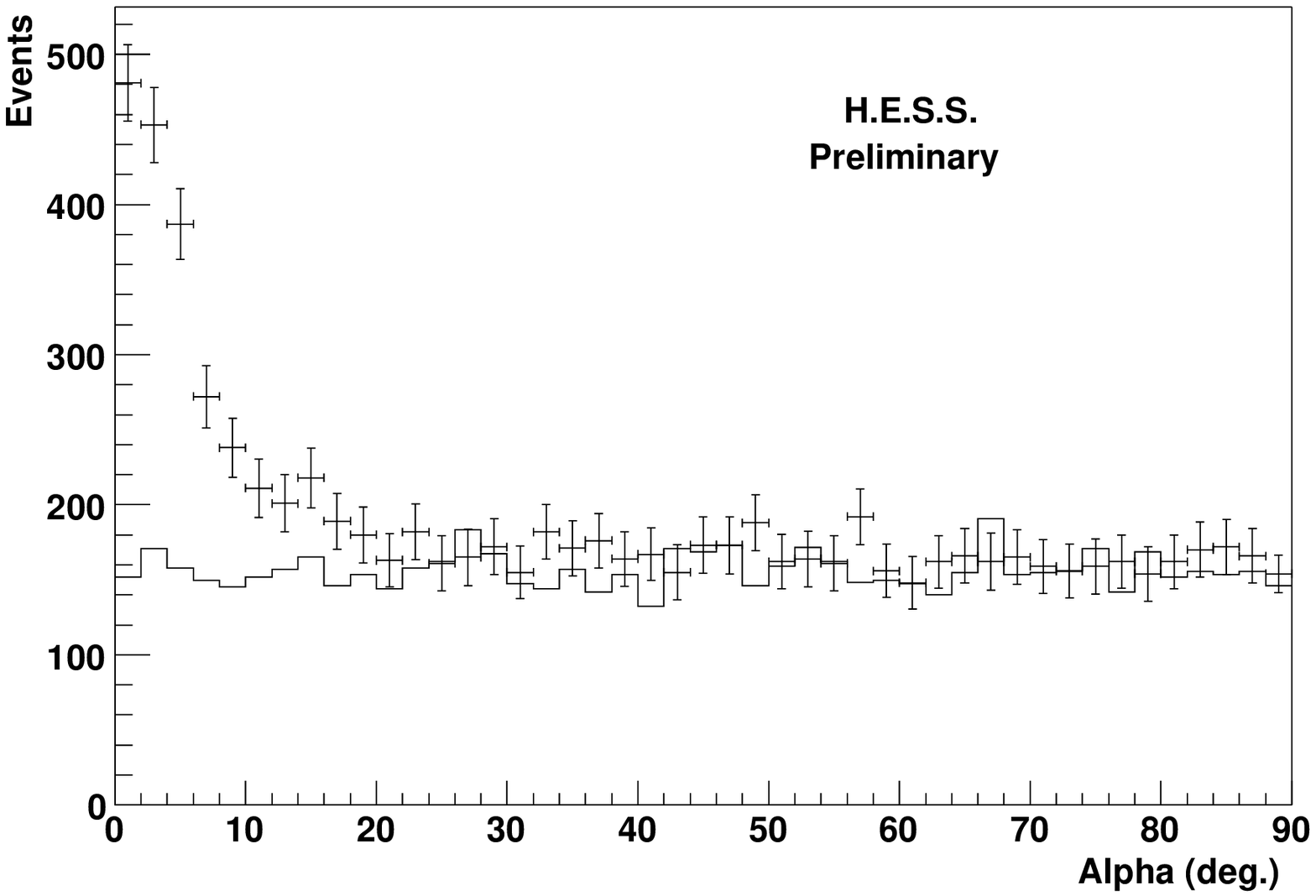}}
\parbox{0.42\linewidth} {\centering \includegraphics[width=13pc]{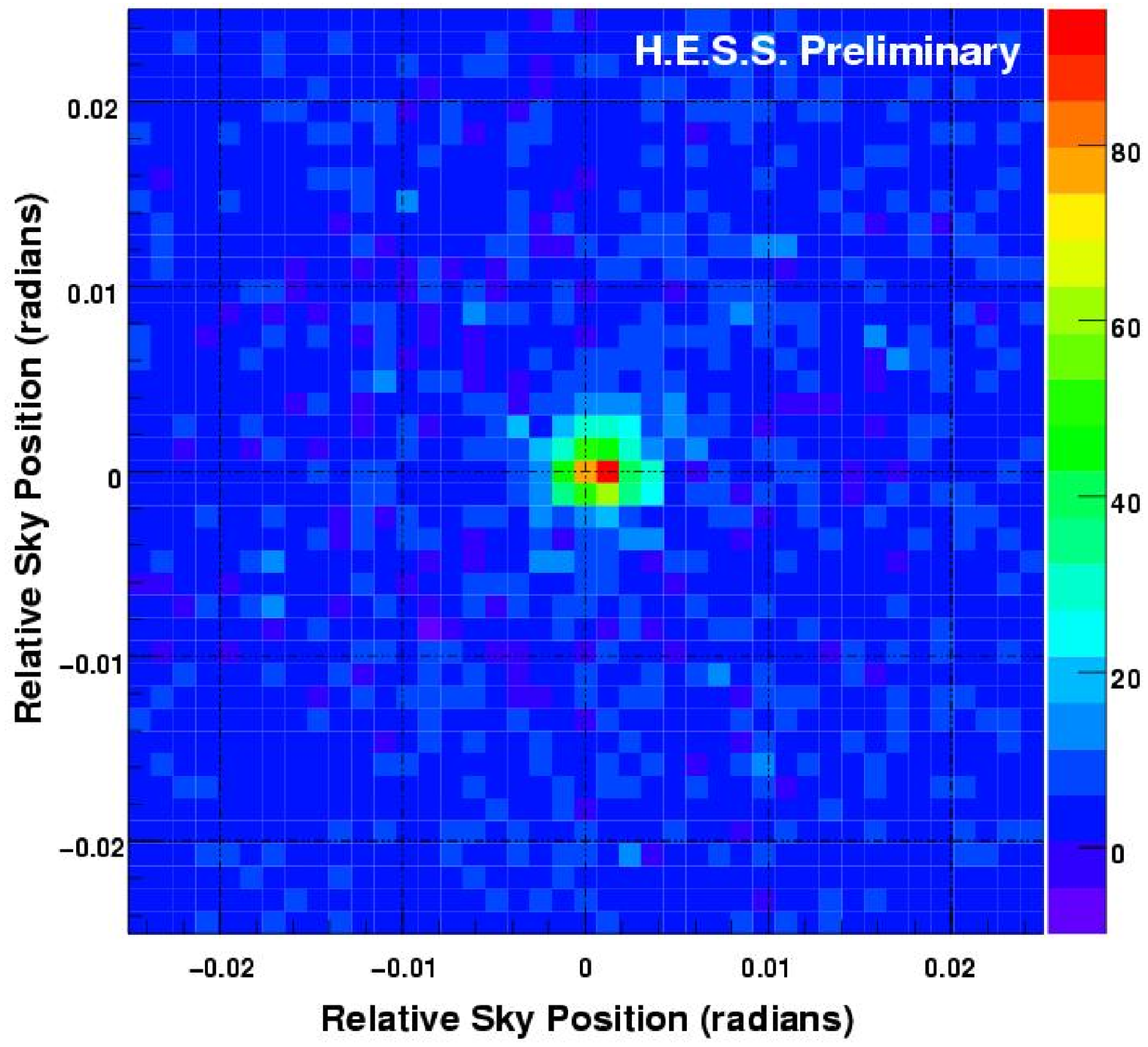}}
\caption{Left: The pointing angle $\alpha$-plot for single telescope Crab 
observations in October 2002. 
Corresponding off-source data (normalised by exposure time) are shown as a histogram.
Right: Reconstructed sky-plot of $\gamma$-ray excess in the region around the Crab.}  
\label{fig_crab}
\end{figure}

The Crab nebula is a standard candle in the VHE $\gamma$-ray field and 
was observed during autumn 2003 to test predictions of the performance
of a single \hess~telescope. Observations were
taken over a zenith angle range of 45$^\circ$\ to 50$^\circ$ during
October and November 2003. A total live-time of 4.7~hours on-source, 
and an equal amount of off-source, data were acquired.
After applying optimised cuts ({\em box cuts} or {\em supercuts}) on
reconstructed Hillas Parameters, a steady rate of 3.6 $\gamma$'s/minute
is observed with a significance of $20.1\sigma$.
Figure~\ref{fig_crab} shows a pointing angle $\alpha$-plot for these data.
Because of the large zenith angle of these observations the post-cut energy threshold
is $\approx$800~GeV.
From Monte Carlo simulations (CORSIKA plus telescope simulation) 
we calculate a preliminary integral 
flux of ($2.64\pm0.20$)$\,\times\,10^{-7}$ m$^{-2}$ s$^{-1}$ ($>1\,$TeV). 

\begin{figure}[h]
\parbox{0.5\linewidth} {\centering \includegraphics[width=15pc]{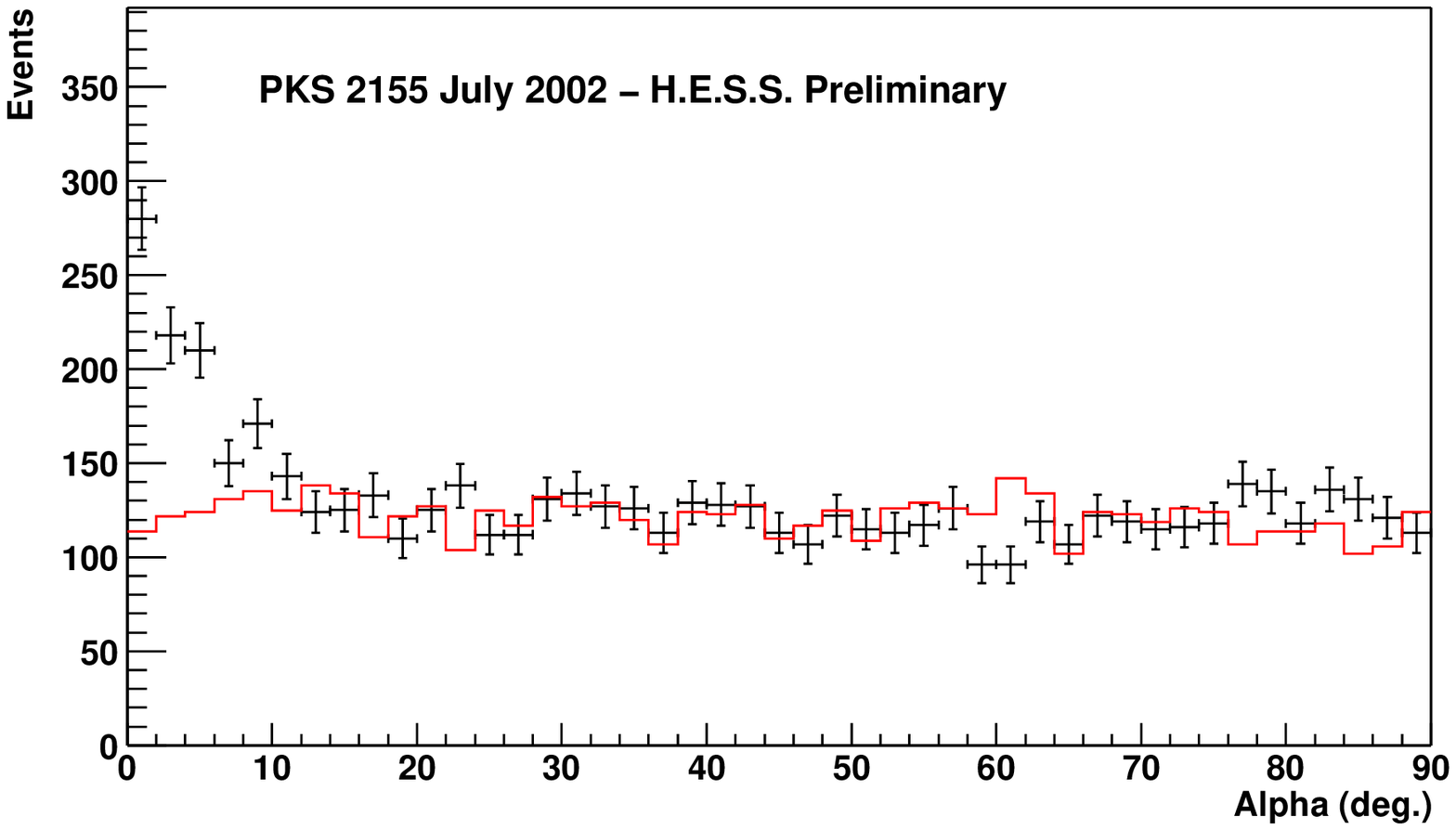}}
\parbox{0.5\linewidth} {\centering \includegraphics[width=15pc]{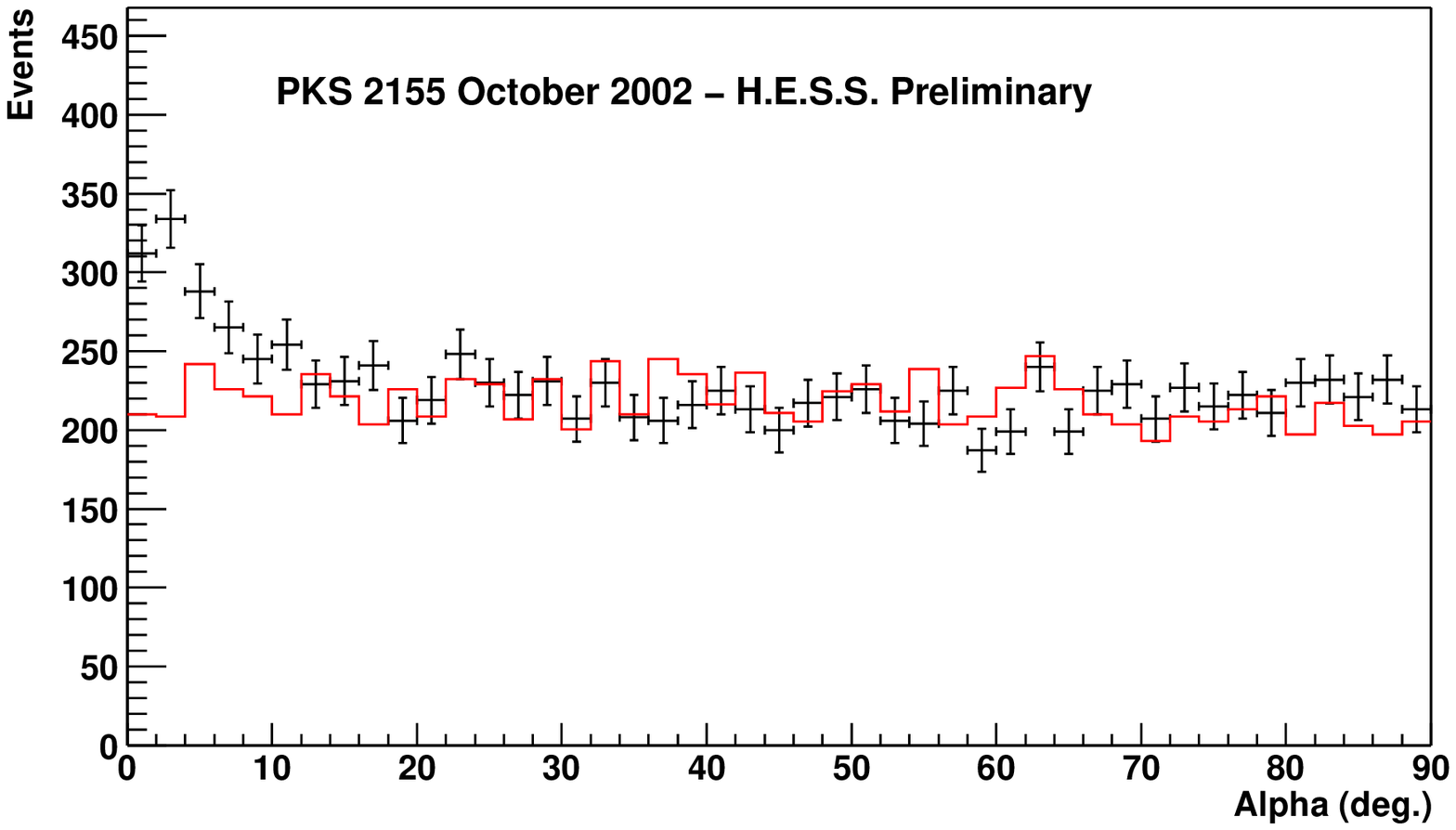}}
\caption{The pointing angle $\alpha$-plot for PKS 2155-304 data from July (left)
 and October (right) 2002. A significant excess is seen in both periods.} 
\label{fig_pks}
\end{figure}

Amongst the targeted extra-galactic objects was the BL-Lac object PKS 2155-304.
Our observations in July and October 2002 confirm the identification of this 
object as a VHE $\gamma$-ray source by the Durham group~\cite{DURHAM_PKS}.
Observations in July with 2.2~hours of live-time on-source show an excess of
$3.1\gamma$'s/minute with a significance of 9.9$\sigma$ (see Fig.~\ref{fig_pks}).
The measured flux in October was significantly lower ($1.2\gamma$'s/minute, 6.6$\sigma$).

\section{First Stereo Observations}

\begin{figure}[h]
\begin{center} 
\includegraphics[width=30pc]{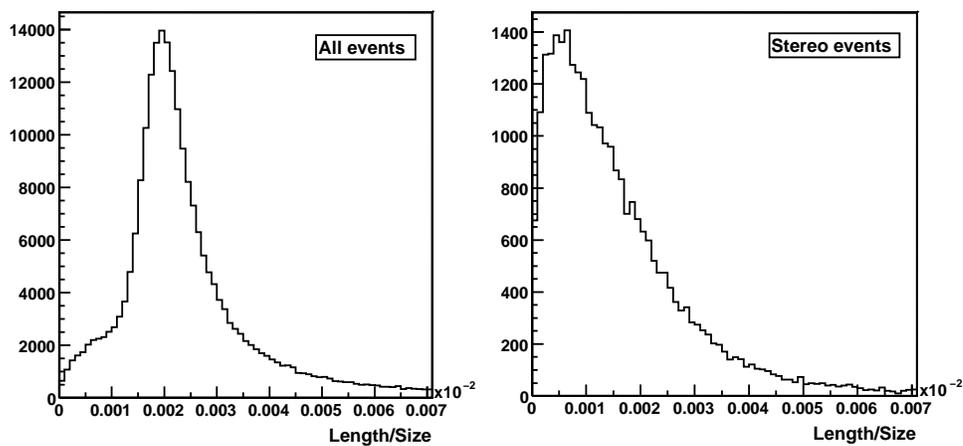}
\end{center}
\caption{Distributions of image {\it Length/Size} for single telescope and coincident events. 
The peak due to single muon images is entirely absent in the stereoscopic data.} 
\label{fig_stereo}
\end{figure}

In March 2003 the first simultaneous observations with two \hess~telescopes were made.
In the initial mode of operation the stereoscopic events are matched in software
using GPS times. Figure~\ref{fig_stereo} shows the distribution of the image {\em 
length/size} for single telescope and coincident events from these data. 
The prominent peak in the
first plot is due to single muon images which are clearly absent in the stereoscopic
data, as expected.

The installation of a hardware central triggering system (expected in June 2003) will allow
us to substantially reduce the camera trigger thresholds whilst maintaining acceptable
levels of dead-time ($<20\%$).

\section{Conclusions}

Two of the four telescopes of \hess~Phase-I are complete and operational. The
optical and mechanical performance of these telescopes meets all specifications.
$\gamma$-ray observations of the Crab confirm the simulated performance of
a single \hess~telescope. We confirm the BL-Lac PKS 2155-304 as a strong VHE 
$\gamma$-ray source.

The full \hess~system should begin operations early in 2004.

\end{document}